\documentclass[11pt]{article}
\pdfoutput=1
\usepackage[a4paper,bottom=4.2cm,top=1cm,head=3cm,width=18cm,dvipdfm]{geometry}
\evensidemargin=\oddsidemargin

\usepackage[dotinlabels]{titletoc}
\usepackage{titlesec}

\usepackage{xcolor}
\usepackage{mathrsfs}
\usepackage{amsmath}
\usepackage{amssymb}
\usepackage{bm}
\usepackage{amsfonts}
\usepackage{subfigure}
\usepackage{feynmp}
\usepackage{extarrows}
\usepackage{slashed}
\usepackage{graphicx}
\usepackage{dcolumn}
\usepackage{verbatim}
\usepackage{here}
\usepackage{multirow} 
\allowdisplaybreaks[4]
\usepackage{indentfirst}
\usepackage{isodateo}

\renewcommand{\thefootnote}{\arabic{footnote}}
\usepackage{isodateo}
\usepackage[CJKbookmarks=true, bookmarksnumbered=true,bookmarksopen=true]{hyperref}
\hypersetup{colorlinks,%
              linkcolor=[rgb]{0,0.2,0.6}, %
              citecolor=[rgb]{0,0.2,0.6}}

\newcommand{\fr}[2]{\mbox{$\frac{\,{#1}\,}{#2}$}}

\def\bge{\begin{equation}}
\def\ede{\end{equation}}
\def\bga{\begin{aligned}}
\def\eda{\end{aligned}}
\newcommand{\beq}{\begin{equation}}
\newcommand{\eeq}{\end{equation}}
\newcommand{\bq}{\begin{equation}}
\newcommand{\eq}{\end{equation}}
\newcommand{\ba}{\begin{array}}
\newcommand{\ea}{\end{array}}
\newcommand{\beqa}{\begin{eqnarray}}
\newcommand{\eeqa}{\end{eqnarray}}
\newcommand{\beqs}{\begin{subequations}}
\newcommand{\eeqs}{\end{subequations}}

\def\({\left(}
\def\){\right)}

\def\leqq{\leqslant}
\def\geqq{\geqslant}
\def\End{\end{document}}

\def\ii{{\tt i}}

\def\ZZ{\mathbb{Z}}

\def\yb{\overline{y}}

\def\End{\end{document}}

\begin{document}

 \thispagestyle{empty}
 \renewcommand{\thefootnote}{\fnsymbol{footnote}}
 \setcounter{footnote}{0}
 \titlelabel{\thetitle.\quad \hspace{-0.8em}}
\titlecontents{section}
              [1.5em]
              {\vspace{4mm} \large \bf}
              {\contentslabel{1em}}
              {\hspace*{-1em}}
              {\titlerule*[.5pc]{.}\contentspage}
\titlecontents{subsection}
              [3.5em]
              {\vspace{2mm}}
              {\contentslabel{1.8em}}
              {\hspace*{.3em}}
              {\titlerule*[.5pc]{.}\contentspage}
\titlecontents{subsubsection}
              [5.5em]
              {\vspace{2mm}}
              {\contentslabel{2.5em}}
              {\hspace*{.3em}}
              {\titlerule*[.5pc]{.}\contentspage}
\titlecontents{appendix}
              [1.5em]
              {\vspace{4mm} \large \bf}
              {\contentslabel{1em}}
              {\hspace*{-1em}}
              {\titlerule*[.5pc]{.}\contentspage}




\begin{center}
{\Large\bf TeV Scale Neutrino Mass Generation, Minimal Inelastic\\
\vspace*{2mm}
           Dark Matter, and High Scale Leptogenesis}

\vspace*{8mm}

{\sc Pei-Hong Gu}\,$^{a}$\footnote{Email: peihong.gu@sjtu.edu.cn}
\,~and~\,
{\sc Hong-Jian He}\,$^{a,b,c,d}$\footnote{Email: hjhe@tsinghua.edu.cn}

\vspace*{4mm}

$^a$\,School of Physics and Astronomy, Shanghai Jiao Tong University, Shanghai 200240, China
\\[1mm]
$^b$\,T.~D.~Lee Institute, Shanghai 200240, China
\\[1mm]
$^c$\,Institute of Modern Physics, Tsinghua University, Beijing 100084, China
\\[1mm]
$^d$\,Center for High Energy Physics, Peking University, Beijing 100871, China

\vspace*{20mm}
\end{center}

\vspace*{3mm}

\begin{abstract}
\baselineskip 17pt
\noindent
The seesaw and leptogenesis commonly depend on the masses of same particles,
and thus are both realized at the same scale. In this work,
we demonstrate a new possibility to realize a TeV-scale neutrino seesaw
and a natural high-scale leptogenesis.
We extend the standard model by two gauge-singlet scalars,
a vector-like iso-doublet fermion and one iso-triplet Higgs scalar.
Our model respects a softly broken
lepton number and an exactly conserved $\ZZ_2^{}$ discrete symmetry.
It can achieve three things altogether:
(i) realizing a testable type-II seesaw
at TeV scale with two nonzero neutrino mass-eigenvalues,
(ii) providing a minimal inelastic dark matter from the new fermion doublets, and
(iii) accommodating a thermal or nonthermal leptogenesis through the singlet scalar decays.
We further analyze the current experimental constraints on our model and discuss the
implications for the dark matter direct detections and the LHC searches.
\\[2mm]
Phys.\ Rev.\ D, in Press [arXiv:1808.09377]
\end{abstract}

\newpage
\renewcommand{\thefootnote}{\arabic{footnote}}
\setcounter{footnote}{0}
\setcounter{page}{2}


\setcounter{footnote}{0}
\renewcommand{\thefootnote}{\arabic{footnote}}

\baselineskip 18pt

\section{Introduction}
\vspace*{1.5mm}
\label{sec:1}

The seesaw\,\cite{minkowski1977,mw1980} extensions of the standard model (SM)
naturally explain the tiny neutrino masses\,\cite{olive2014}, while accommodating
a leptogenesis\,\cite{fy1986,lpy1986,LG-Rev} mechanism to generate the observed
cosmic baryon asymmetry\,\cite{olive2014}.
In the conventional seesaw-leptogenesis scenarios,
the scales of generating neutrino masses and baryon asymmetry are tied together,
and determined by the masses of the same particles. The leptogenesis could not be
realized at the TeV scale unless it invokes a large fine-tuning to resonantly enhance
the required CP asymmetry. This means that a natural leptogenesis is achieved at
high scale, and the conventional scenarios link the seesaw to the same
leptogenesis scale, which prevent the realization of testable seesaw at the TeV scale.

\vspace*{1mm}

The strong evidence for non-baryonic dark matter (DM) poses another great challenge
to the modern particle physics and cosmology\,\cite{olive2014}. There have been interesting ideas
explaining the DM puzzle. For instance, the minimal DM models \cite{cfs2006,cst2007,chpst2015}
can give testable predictions for DM properties
including the DM mass and the DM-nucleon scattering.
However, for models with the new weak multiplet of nonzero hypercharge, its neutral DM component
will have gauge interactions with $Z^0$, and thus is excluded by the
direct DM searches\,\cite{cfs2006}.  This calls for viable extensions.
Besides, the DM particle may also play an important role in the generation of neutrino masses
\cite{knt2003,ma2006,typeII2009,lugu2015,chsvv2017}
and the realization of baryon asymmetry \cite{typeII2009}.

\vspace*{1mm}

In this work, we propose an attractive possibility
that new physics for generating a testable TeV-scale seesaw can accommodate
a thermal or inflationary baryogenesis at a very high scale.
At the same time, we provide a viable minimal inelastic DM candidate at the TeV scale.
In our construction, we will construct a realistic model including two gauge-singlet scalars,
a vector-like iso-doublet fermion and one iso-triplet Higgs scalar besides the SM fields.
Our model has a softly broken lepton number and an exactly conserved $\ZZ_2^{}$ discrete symmetry,
so it differs from other models\,\cite{gu2017} with a spontaneous breaking lepton or baryon number.
Under such softly broken lepton number and the exact $\ZZ_2^{}$ symmetry,
our model can achieve three things altogether:
(i) realizing a testable type-II seesaw at TeV scale
with two nonzero neutrino mass-eigenvalues,
(ii) providing a minimal inelastic dark matter from the new fermion doublets,
with the mass-splitting induced by interactions related to the neutrino mass-generations,
and (iii) accommodating a thermal or inflationary leptogenesis at high scale
through the scalar-singlet decays.
Although the leptogenesis scale is high, realizing the DM relic density in
our scenario requires the DM mass to be about $1.2$\,TeV.
As we will show, the present minimal inelastic DM is a stable Majorana fermion, and
depends on two new parameters:
the DM mass and the mass difference between the DM and another particle.
(This differs from the previous minimal DM model\,\cite{cfs2006,cst2007,chpst2015}
where the DM is either a scalar or Dirac fermion,
and its tree-level mass is the only new physics parameter.)
The predicted Higgs triplet and DM fermion of our model can be searched at the LHC and
future high energy $pp$ colliders. The same DM particle can be probed
by the direct and indirect DM detection experiments\,\cite{DM-DirectExp-Rev}.

\vspace*{1mm}

This paper is organized as follows.
In Section\,\ref{sec:2}, we present the model setup.
Then, we study the minimal inelastic DM in Section\,\ref{sec:3}
and the radiative type-II neutrino seesaw in Section\,\ref{sec:4}.
The realization of high scale leptogenesis is presented Section\,\ref{sec:5}.
Finally, we conclude in Section\,\ref{sec:6}.

\section{Model Construction}
\vspace*{1.5mm}
\label{sec:2}

In this section, we present the model setup and discuss the involved (un)broken symmetries.
For the current model, we introduce two gauge-singlet scalars,
\begin{eqnarray}
\sigma_{j}^{}(1,1,0)=\frac{1}{\sqrt{2}\,}(\sigma_{jR}^{}+\ii\sigma_{jI}^{}),
~~~(j=1,2),
\end{eqnarray}
a vector-like iso-doublet fermion (with left-handed component $\psi_L^{}$
and right-handed component ${\psi_L'}^{c}$\,),
\begin{eqnarray}
\psi^{}_L\!\(1,2,+\fr{1}{2}\!\)
=\left(\!\!\begin{array}{c} \xi^{+}_{L} \\[2mm]
\chi^{}_{L} \end{array}\!\!\right)\!,~~~
\psi^{\prime}_L\!\(1,2,-\fr{1}{2}\!\)
=\(\!\!\begin{array}{c} \chi'^{}_{L} \\[2mm]
\xi'^{-}_{L} \end{array}\!\!\!\)\!,
\end{eqnarray}
and an iso-triplet scalar,
\begin{eqnarray}
&&\Delta(1,3,-1)=\left\lgroup\!\!
\begin{array}{cc}
\fr{1}{\sqrt{2}\,}\delta^{-}_{} & \delta^{0}_{}
\\[2.5mm]
\delta^{--}_{} & -\fr{1}{\sqrt{2}\,}\delta^{-}_{}
\end{array}
\!\!\right\rgroup\!.
\hspace*{10mm}
\end{eqnarray}
On the left-hand-side of each equation above,
the numbers in the parentheses describe the representations (or quantum numbers)
of the corresponding field under the SM gauge group
$SU(3)_c^{}\otimes SU(2)_L^{}\otimes U(1)^{}_{Y}$.\,
This model also respects a softly broken lepton number and an exactly conserved $\ZZ_2^{}$
discrete symmetry. By definition, only the scalar singlets $\sigma_j^{}$ ($j=1,2$)
carry a lepton number $-1$, which is opposite to the SM leptons.
Under the $\ZZ_2^{}$ discrete symmetry, the SM fields and the scalar triplet
\,$\Delta$\,
are $\ZZ_2^{}$ even, while the scalar singlets $\,\sigma_j^{}\,$ and the fermion doublets
\,($\psi^{}_{L},\,\psi'^{}_{L})$\, are $\ZZ_2^{}$ odd.
Thus, we have the following $\ZZ_2^{}$ transformations,
\begin{eqnarray}
(\textrm{SM},~\Delta)&\stackrel{\ZZ_2^{}}{\longrightarrow}&
(\textrm{SM},~\Delta)\,,
\nonumber\\
(\sigma_j^{},~\psi^{}_L,~\psi'^{}_L)
&\stackrel{\ZZ_2^{}}{\longrightarrow}&
-(\sigma_j^{},~\psi^{}_L,~\psi'^{}_L)\,.
\end{eqnarray}
Since the $\ZZ_2^{}$ discrete symmetry is exactly conserved,
the scalar singlets $\sigma_j^{}$ will not acquire any non-zero
vacuum expectation value (VEV).

\vspace*{1mm}

For the current analysis, we write down the following relevant Lagrangian terms,
\begin{eqnarray}
\label{eq:L}
\label{lag}
\hspace*{-3mm}
\mathcal{L} &\!\supset\!\!&
-\sigma_{}^\dagger M_\sigma^2\sigma\!
-\fr{1}{2}\sigma^T\!\tilde{M}^2_\sigma\sigma\!
-M_\Delta^2\textrm{Tr}(\Delta^\dagger_{}\Delta)
+\mu_{\Delta\phi}^{}\phi^T_{}\!\ii\tau_2^{}\Delta\phi
\nonumber\\[1mm]
\hspace*{-3mm}
&&
-M_\psi^{}\overline{\psi_L^{c}}\ii\tau_2^{}\psi'^{}_{L}
+\fr{1}{2}f\overline{\psi_{L}^{c}}\ii\tau_2^{}\Delta \psi_L^{}
-\fr{1}{2}f'\overline{\psi'^{c}_{L}}\ii\tau_2^{}\Delta^\dagger_{} \psi'^{}_L
\nonumber\\[1mm]
\hspace*{-3mm}
&&
-y^{}_{\alpha j}\overline{L_{L\alpha}^c} \ii\tau_2^{} \psi_L^{}\sigma_j^{}
+\textrm{H.c.},
\end{eqnarray}
where $\tau_2^{}$ is the second Pauli matrix and
$\,\sigma =(\sigma_1^{},\,\sigma_2^{})^T\,$ denotes the singlet scalars.\,
In Eq.\eqref{lag}, the fields $\phi$, $L_{L}^{}$, and $e_R^{}$ denote the Higgs doublet,
the left-handed lepton doublet, and the right-handed lepton in the SM, respectively.
Thus, the Higgs doublet $\phi$ and the left-handed lepton doublet $L_{L}^{}$
take the following form,
\begin{eqnarray}
\phi\!\(1,2,+\fr{1}{2}\!\)=\!\(\!\!\begin{array}{c} \phi^{+}_{}
\\[2mm]
\phi^{0}_{}\end{array}\!\!\)\!,~~~~
L_{L}^{}\!\!\(1,2,-\fr{1}{2}\!\)=\!\(\!\!\begin{array}{c} \nu^{}_{L} \\[2mm]
e_{L}^{}\end{array}\!\!\)\!.
\end{eqnarray}
It is clear that in Eq.(\ref{lag})
the $\tilde{M}^2_\sigma$ mass term is the unique source of the lepton number violation,
since by construction only the SM leptons and the scalar singlets $\sigma_j^{}$
carry lepton numbers.
We note that requiring the softly broken lepton number and the exactly conserved $\ZZ_2^{}$
discrete symmetry has forbidden the following gauge-invariant terms,
\begin{eqnarray}
\hspace*{-2mm}
\mathcal{L} \hspace*{-1mm}&\supset\!\!\!\!\!\!/&\!
-M'\overline{\psi^{c}_L} \ii\tau_2^{} L^{}_L\!
-\fr{1}{2}f''\overline{\psi'^c_{L}} \ii\tau_2^{}\Delta^\dagger_{}\! L^{}_{L}\!
-\fr{1}{2}f'''\overline{L^c_{L}} \ii\tau_2^{}\Delta^\dagger_{}\! L^{}_{L}
-f''''_j\bar{L}_L^{}\psi'_L\sigma_j^{}
-y'\overline{\psi'^{}_{L}}\phi\, e_{R}^{}+\textrm{H.c.}\,,
\hspace*{8mm}
\end{eqnarray}
with $e_R^{}(1,1,-1)$ being the right-handed leptons.

\vspace*{1mm}

For simplicity, we consider the real mass parameters
in the present analysis,
\begin{eqnarray}
\mu_{\Delta\phi}^{}=\mu_{\Delta\phi}^{\ast}\,,~~~
M_\psi^{}=M_{\psi}^{\ast}\,.
\end{eqnarray}
Then, we choose the singlet mass matrix $M_{\sigma}^2$
to be diagonal, and for simplicity of the analysis we further assume
$\tilde{M}_{\sigma}^2$ to be diagonal as well,
\begin{eqnarray}
M_{\sigma}^2=\textrm{diag}\{M_{\sigma_{1}^{}}^2,\,M_{\sigma_{2}^{}}^2\},~~~
\tilde{M}_{\sigma}^2
=\textrm{diag}\{\tilde{M}_{\sigma_{1}^{}}^2,\,\tilde{M}_{\sigma_{2}^{}}^2\}.
\end{eqnarray}
Accordingly, we can deduce the mass eigenvalues of the real and imaginary components
of the two singlet scalars $\sigma_{1,2}^{}$ as follows,
\beqs
\begin{eqnarray}
&& \hspace*{-10mm}
\mathcal{L}_{\text{mass}}^{\sigma}
\,=\, -\frac{1}{2}M_{\sigma_{jR}^{}}^2 \sigma_{jR}^2
-\frac{1}{2}M_{\sigma_{jI}^{}}^2 \sigma_{iI}^2,
\\[1.5mm]
&& \hspace*{-10mm}
M_{\sigma_{jR}^{}}^2\!=M_{\sigma_{j}^{}}^2\!+\tilde{M}_{\sigma_{j}^{}}^2,~~~
M_{\sigma_{jI}^{}}^2\!=M_{\sigma_{j}^{}}^2\!-\tilde{M}_{\sigma_{j}^{}}^2.~~~~~~~
\end{eqnarray}
\eeqs
For simplicity of demonstration, we also set both Yukawa couplings $f$ and $f'$ be real,
\begin{eqnarray}
f=f^\ast_{}\,,~~~~~f'^{}=f'^\ast_{}\,,
\end{eqnarray}
although one of them is allowed to be complex.

\vspace{2mm}
\section{Minimal Inelastic Dark Matter}
\label{sec:3}

In this section, we further analyze the model predictions and the
experimental constraints. In particular, we shall identify
a stable Majorana fermion as a viable DM particle of mass $\sim\!1.2$\,TeV.
We show that the present inelastic DM model depends on two new parameters:
the DM mass and the mass difference between the DM and another particle.
This differs from the previous minimal DM model\,\cite{cfs2006,cst2007,chpst2015}
where the DM is either a scalar or Dirac fermion,
and its tree-level mass is the only new physics parameter.

In the present model, the SM Higgs doublet $\phi$ will develop a VEV
for spontaneous electroweak symmetry breaking at the weak scale.
The scalar triplet $\Delta$ has a positive mass-term and will acquire an induced VEV,
due to its cubic interaction with the SM Higgs doublet $\phi$\, via the
$\mu_{\Delta\phi}^{}$ term in Eq.(\ref{lag}).
So we will refer the scalar triplet $\Delta$ as a Higgs triplet.
The Higgs scalars $\phi$ and $\Delta$ have 10 degrees of freedom in total,
including 4 real neutral scalars, 2 singly charged scalars,
and 1 doubly charged scalar,
\beqs
\begin{eqnarray}
\phi&=& \!\!\left\lgroup\!\!
\begin{array}{c}
\phi^{+}_{}
\\[2mm]
\fr{1}{\sqrt{2}\,}(v_\phi^{}\!+h_\phi^{}\!+\ii\phi^{0}_I)
\end{array}\!\!\right\rgroup \!\!,
\\[2mm]
\Delta&=& \!\!\left\lgroup\!\!
\begin{array}{cc}
\fr{1}{\sqrt{2}\,}\delta^{-}_{} ~&~
\fr{1}{\sqrt{2}\,}(v_\Delta^{}\!\!+\!h_\Delta^{}\!\!+\ii\delta_I^{0})
\\[2mm]
\delta^{--}_{} & -\fr{1}{\sqrt{2}\,}\delta^{-}_{}
\end{array}\!\!\right\rgroup \!\!,
\end{eqnarray}
\eeqs
with the VEVs $\,v_{\phi}^{}\,$ and $\,v_\Delta^{}\,$ from their neutral components.
Among the degrees of freedom in the Higgs scalars $\phi$ and $\Delta$,
one neutral massless eigenstate from $\,\phi_I^0\,$ and $\,\delta_I^0\,$ as well as
one pair of charged massless eigenstates from $\,\phi^{\pm}_{}\,$ and $\,\delta^{\pm}_{}\,$
will be absorbed by the longitudinal components
of the weak gauge bosons $Z^0$ and $W^\pm$.\,
The VEVs $v_{\phi}^{}$ and $v_{\Delta}^{}$ should subject to the
precision constraints \cite{olive2014},
\begin{eqnarray}
v&\equiv&\sqrt{ v_\phi^2+2v_\Delta^2\,} \,\simeq\, 246\,\textrm{GeV}\,,
\nonumber\\[-5mm]
\\
\rho &=&\frac{\,v_\phi^2\!+2v_\Delta^2\,}{\,v_\phi^2\!+4v_\Delta^2\,}
\,=\, 1.00040\pm 0.00024\,.
\nonumber
\end{eqnarray}
With the $3\sigma$ lower limit $\,\rho\geqq 0.99968\,$,\, we deduce
\begin{eqnarray}
\label{htvev}
|v_\Delta^{}|=\sqrt{\frac{\,1\!-\!\rho\,}{2\rho\,}}v
\leqq 3.1\,\textrm{GeV} .
\end{eqnarray}
From the relevant scalar potential terms in Eq.\eqref{lag},
we derive $v_{\Delta}^{}$ as follows,
\begin{eqnarray}
\label{eq:v_Delta}
v_\Delta^{} \,\simeq\,
\frac{~\mu_{\Delta\phi}^{}v_\phi^2~}
{~\sqrt{2}\,M_\Delta^2~}\,,
\end{eqnarray}
for $\,M_\Delta^{}\gg \mu_{\Delta\phi}^{}$\,,\,
or,\, $M_\Delta^{}\gg v_\phi^{}$\,.\,
We see that a small triplet VEV $\,v_{\Delta}^{}$\, is naturally generated in the
present model, due to the seesaw-type suppression in the above formula \eqref{eq:v_Delta}.

\vspace*{1mm}

According to the Yukawa interactions in Eq.\eqref{lag},
the small VEV $v_\Delta^{}$ of Higgs triplet
will also contribute to Majorana masses of the neutral fermions
$\chi_L^{}$ and $\chi'^{}_L$\,.\, Thus, we derive the following mass terms
for charged fermions $(\xi^{\pm}_L,\,{\xi'}^{\pm}_L)$ and
neutral fermions $(\chi_L^{},\,{\chi'}_L^{})$,
\begin{eqnarray}
\label{eq:chi-xi-mass}
\mathcal{L}^{\xi\chi}_{\text{mass}}
\,=\, -M_\psi^{}\overline{\xi^{+c}_{L}}\xi'^{-}_L
+M_\psi^{}\overline{\chi^{c}_L}\chi'^{}_L
    -\frac{v_\Delta^{}}{2\sqrt{2}\,}\left(f\overline{\chi^{c}_L}\chi^{}_L
    +f'\overline{\chi'^{c}_L}\chi'^{}_L\right) + \textrm{H.c.}
\end{eqnarray}
For convenience, we can express the Dirac spinors in terms of the left-handed
Weyl spinors in the $(\frac{1}{2},\,0)$ representation of Lorentz group,
\beqa
\hspace*{-3mm}
\chi_L  \!=\!\(\!\!\ba{c} \chi \\[1mm] 0 \ea\!\!\)\!,~~
\chi'_L \!=\!\(\!\!\ba{c} \chi' \\[1mm] 0 \ea\!\!\)\!,~~
\xi_L^{\pm}\!=\!\(\!\!\ba{c} \xi^{\pm} \\[1mm] 0 \ea\!\!\)\!,~~
\xi_L^{\prime\pm}\!=\!\(\!\!\ba{c} \xi^{\prime\pm} \\[1mm] 0 \ea\!\!\)\!.~~
\eeqa
Thus, we can rewrite the mass terms \eqref{eq:chi-xi-mass} as follows,
\beqa
\label{eq:chi-xi-mass2}
\mathcal{L}^{\xi\chi}_{\text{mass}} \,=\,
M_\psi^{} \xi^{+T}\!\epsilon\,{\xi'}^-\!\!
-M_\psi^{}\chi^T\!\epsilon\chi'
+\frac{v_\Delta^{}}{2\sqrt{2}\,}\!
\( f\chi^T\!\epsilon\chi +
   f'\chi^{\prime T}\!\epsilon\chi'\)
+ \textrm{H.c.},~~~~~
\eeqa
where $\,\epsilon = \ii\tau_2^{}\,$ is anti-symmetric.
We see that the two charged Weyl spinors $\xi^\pm$ and $\xi^{\prime\pm}$ form a
Dirac mass term with mass $\,M_\xi^{}=M_\psi^{}$.\,
Defining the charged Dirac spinor,
\beqa
\tilde{\xi}^\pm =
\(\! \ba{c}
\xi^\pm \\[1mm]
\epsilon{\xi^{\prime\mp}}^*
\ea\!\!\)
=\, \xi_L^\pm + ({\xi_L^{\prime\mp}})^c
,
\eeqa
we can express the Dirac mass term
$\,M_\psi^{} \xi^{+T}\!\epsilon\,{\xi'}^- \!\!+\textrm{H.c.}$
in the conventional 4-component form
$M_\psi^{} \overline{\tilde{\xi}^+}\,{\tilde\xi^+}$\,.\,
From Eq.\eqref{eq:chi-xi-mass2}, the neutral fermions
$\,\widehat{\chi}\equiv (\chi^{},\,\chi'^{})^T$
have the Majorana mass-term
$\,-\fr{1}{2}\widehat{\chi}^T\mathbb{M}_\chi^{}\widehat{\chi}\,$
with mass matrix,
\beqa
\label{eq:MF}
\mathbb{M}_\chi^{}=\(\!\!\ba{cc}
-f\bar{v}_\Delta^{} & M_\psi^{} \\[2mm]
M_\psi^{} & -f'\bar{v}_\Delta^{}
\ea\!\!\)\!,
\eeqa
where $\,\bar{v}_\Delta^{}=v_\Delta^{}/\!\sqrt{2}$\,.\,
Then, we can diagonalize the symmetric mass matrix
$\mathbb{M}_\chi^{}$ and derive the following mass-eigenvalues,
\beqs
\beqa
\hspace*{-14mm}
&&\!M_{\chi_1^{}}^{}\!
=\sqrt{M_\psi^2+\!\frac{\,v_\Delta^2}{8}(f\!-\!f')^2_{}\,}
-\frac{\,f\!+\!f'\,}{2\sqrt{2}}v_\Delta^{},
\\[2mm]
\hspace*{-14mm}
&&\!M_{\chi_2^{}}^{}\!
=\sqrt{M_\psi^2+\!\frac{\,v_\Delta^2}{8}(f\!-\!f')^2_{}\,}
+\frac{\,f\!+\!f'\,}{2\sqrt{2}}v_\Delta^{},
\end{eqnarray}
\eeqs
with $\,M_{\chi_1^{}}^{}<M_{\chi_2^{}}^{}\,$
for $f\!+\!f'>0$.\,
For the case
$\,M_\psi^{}\gg |f\!\pm\!f'|\,v_\Delta^{}\!=\mathcal{O}(\text{GeV})\,$,\,
we see that the mass-eigenvalues $\,M_{\chi_1^{}}^{}$ and
$\,M_{\chi_2^{}}^{}$ are quite degenerate:
$\,M_{\chi_1^{}}^{}\approx M_{\chi_2^{}}^{}\approx M_\psi^{}$.\,
For diagonalizing the mass matrix \eqref{eq:MF},
we rotate the fields $(\chi_1^{},\,\chi_2^{})$
into their mass-eigenstates $(\widetilde\chi_1^{},\,\widetilde\chi_2^{})$
by the unitary rotation
$(\chi,\,\chi')^T=U(\chi_1^{},\,\ii\chi_2^{})^T$
with
\beqa
\label{eq:U}
U =\(\!\!\!\ba{rc}
 \cos\theta ~&~ \sin\theta
\\[2mm]
-\sin\theta ~&~ \cos\theta
\ea
\!\)\!.
\eeqa
Thus, we determine the rotation angle $\theta$ as follows,
\begin{eqnarray}
\tan 2\theta \,=\,
\frac{2\sqrt{2}M_\psi^{}}{~(f\!-\!f')v_\Delta^{}~}\,.
\end{eqnarray}
For the case $\,M_\psi^{}\gg |f\!-\!f'|v_\Delta^{}=\mathcal{O}(\text{GeV})$,\,
we have the rotation angle $\,\theta \simeq \fr{\pi}{4}$.\,

\vspace*{1mm}

With these, we can rewrite the mass term \eqref{eq:chi-xi-mass2} in the
diagonalized form,
\beqa
\label{eq:chi-xi-massD}
\mathcal{L}^{\xi\chi}_{\text{mass}} \,=\,
M_\psi^{}\overline{\tilde\xi^+}{\tilde\xi^+}\!\!
-\fr{1}{2}\!
\!\(\! M_{\chi_1^{}}^{}\!\chi_1^T\!\epsilon\chi_1^{} \!+\!
   M_{\chi_2^{}}^{}\chi_2^T\!\epsilon\chi_2^{}\!\)\!
+ \textrm{H.c.}~~~~~~~
\eeqa
Then, we can derive gauge interactions of
the charged Dirac fermions $\,\tilde\xi^{\pm}_{}\,$
and the neutral Majorana fermions $\,\chi_{1,2}^{}\,$.\,
In the usual 4-component notations, we denote
$\,\chi_{Lj}^{}=(\chi_j^{},\,0)^T\,$ ($j=1,2$). Thus, we can
express the gauge interactions of $\,\tilde\xi^{\pm}_{}\,$
and $\,\chi_{L1,2}^{}$ as follows,
\begin{eqnarray}
\mathcal{L}^{\xi\chi}_{\text{G}} &=&
e\,\overline{\tilde\xi^+_{}}\gamma^\mu_{}\tilde\xi^+_{}A_\mu^{}\!
+\frac{\,g\cos 2\theta_W^{}\,}{2\cos\theta_W^{}}
\overline{\tilde\xi^+_{}}\gamma^\mu_{}\tilde\xi^+_{}Z_\mu^{}
+\,\frac{g}{\,2\cos\theta_W^{}\,}
\overline{\chi_{L1}^{}}\gamma^\mu_{}\chi_{L2}^{}Z_\mu^{}
\nonumber\\
&&
-\,\frac{\,g\,}{2}\overline{\chi_{L1}^{}}\gamma^\mu_{}\tilde\xi^{+}_{}W_\mu^{-}
-\frac{\,g\,}{2}
\overline{\chi_{L2}^{}}\gamma^\mu_{}\tilde\xi^{+}_{}W_\mu^{-}\!
+\textrm{H.c.},
\end{eqnarray}
where we have set $\,\theta \simeq \frac{\pi}{4}\,$,\,
which holds well for $\,M_\psi^{}\gg(f\!+\!f')v_\Delta^{}$.\,

\vspace*{1mm}

As Eq.(\ref{htvev}) restricts the Higgs triplet VEV $\,v_\Delta^{}$\, within a few GeV,
we find that the mass-splitting between the Majorana fermions
$\,(\chi_1^{},\,\chi_2^{})$\, is constrained as follows,
\begin{eqnarray}
\label{mmsplit}
\hspace*{-4mm}
\Delta M_{\chi}^{} &\!\!=\!&
M_{\chi_2^{}}\!-M_{\chi_1^{}}=\frac{1}{\sqrt{2}\,}(f\!+\!f')v_\Delta^{}
\nonumber\\
\hspace*{-4mm}
&\!\!=\!&
 17.1\,\textrm{GeV}\(\!\frac{f\!+\!f'}{\,2\sqrt{4\pi}~}\!\)
\(\!\frac{v_\Delta^{}}{\,3.1\,\textrm{GeV}~}\)
\lesssim 17.1\,\textrm{GeV},
\end{eqnarray}
which is much smaller than the $\chi_{1,2}^{}$ masses themselves.
This means that the Majorana fermions $\chi_{1,2}^{}$ are quasi-degenerate
and thus are pseudo-Dirac fermions.

\vspace*{1mm}

In the next section, we shall show that the Higgs triplet VEV $v_\Delta^{}$
is also responsible for the neutrino mass-generations through
a radiative type-II seesaw at TeV scale.
Such TeV-scale Higgs triplet can be tested at the LHC.
The LHC could further probe the structure of the neutrino mass matrix if the VEV
of this Higgs triplet is not bigger than $10^{-4}_{}$\,GeV \cite{fhhlw2008}.
In this case, the mass-splitting (\ref{mmsplit}) should receive an upper bound,
\begin{eqnarray}
\label{mmsplit2}
\Delta M_{\chi}^{}~\lesssim~ 550\,\textrm{keV},
\end{eqnarray}
for $\,|f|,|f'|<\sqrt{4\pi}\,$ and $\,v_\Delta^{}\lesssim 10^{-4}_{}\,$GeV.\,
On the other hand, even if the Majorana fermions $\chi_{1,2}^{}$
are degenerate and thus compose a Dirac fermion
$\,\chi=\chi_L^{}+\chi'^{c}_L$,\,
the radiative corrections from the electroweak gauge-interactions
will induce a mass-splitting between the charged fermions $\,\tilde\xi^{\pm}_{}$
and the neutral fermion $\chi$\,,
\begin{eqnarray}
\label{eq:DM-xi-chi}
\Delta M \,=\, M_{\tilde\xi^{\pm}_{}}^{}\!-M_{\chi^{}}^{}\!
\,=\, \frac{\,g^2_{}\!\sin^2_{}\!\theta_W^{}\,}{16\pi^2_{}}M_\psi^{}
 F\!\(\!\frac{m_Z^{}}{M_\psi}\!\) \!,~~~~~~
\end{eqnarray}
where $\theta_W^{}$ denotes the weak mixing angle and
$m_Z^{}$ is the mass of gauge boson $Z^0$.\,
The function $F$ is defined as
\begin{eqnarray}
\!\!F(r)\!\!&=&\!\!\!
\left\{\,\begin{array}{lll}\!\!\!
r^4_{}\ln r \!-\!r^2_{}\!-\!r(r^2\!-\!4)^{\frac{1}{2}}_{}(r^2_{}\!+\!2)
\displaystyle \ln \frac{\,r\!+\!\sqrt{r^2\!-\!4}\,}{2},
\hspace*{5mm}
(\textrm{for}~r\geqq 2);
\\[3mm]
\!\!\!r^4_{}\ln r\!-\!r^2_{}\!+\!r(4\!-\!r^2)^{\frac{1}{2}}_{}(r^2_{}\!+\!2)
\displaystyle
\arctan\frac{\sqrt{4\!-\!r^2}}{r} ,
\hspace*{5mm}
(\textrm{for}~r\leqq 2).
\ea
\right.
\end{eqnarray}
For $\,M_\psi^{}\gg m_{Z}^{}$, the radiative mass-splitting
$\,\Delta M\,$ is well approximated as
\begin{eqnarray}
\label{cmsplit}
\Delta M \,\simeq\, \frac{1}{2}\alpha_{\text{em}}^{} m_Z^{}
\,\simeq\, 356\,\textrm{MeV} ,
\end{eqnarray}
where $\,\alpha_{\text{em}}^{}=e^2/4\pi\simeq 1/128\,$ is the fine structure constant
at the scale $\,\mu =m_Z^{}\,$.\,
If the tree-level mass-splitting (\ref{mmsplit}) is smaller than
the radiative mass-splitting (\ref{cmsplit}), such as the choice (\ref{mmsplit2}),
the charged fermions $\tilde\xi^{\pm}_{}$ can decay into the neutral fermion $\chi$
with a virtual $W^{\pm}_{}$ boson.
Subsequently, the heavier Majorana fermion $\chi_{2}^{}$ can decay
into the lighter Majorana fermion $\chi_{1}^{}$ with a virtual $Z$ boson.
Alternatively, if the mass-splitting (\ref{mmsplit}) is larger than
the mass-splitting (\ref{cmsplit}), the heavier $\chi_{2}^{}$ can
simultaneously decay into $\tilde\xi^{\pm}_{}$ and $W^*$,\,
and then $\tilde\xi^{\pm}_{}$ will decay into the lighter $\chi_1^{}$ plus $W^*$.\,
In either case, all of the decay chains can be completed
before the Big Bang Nucleosynthesis (BBN) epoch.
For instance, we may consider the loop-induced process
$\,\chi_2^{}\!\to\! \chi_1^{}\gamma\,$,\, which always exists [for any
mass-difference in \eqref{mmsplit} or \eqref{eq:DM-xi-chi}], and whose decay rate may be
slower than (or comparable to) the tree-level processes mentioned above.
If this loop-induced decay process is much faster than the BBN epoch,
then the above analysis is fine.
We note that the $\chi_1^{}$-$\chi_2^{}$-$\gamma$ vertex can arise
from the dimension-5 effective operator,
\begin{eqnarray}
\mathcal{L}\supset
- \frac{i}{\Lambda} \bar{\chi}_{1}^{}\sigma^{\mu\nu}_{}\!\chi_2^{} A_{\mu\nu} \,,
\hspace*{10mm}
\textrm{with}~~\frac{1}{\Lambda}\sim
\frac{\(\frac{g}{2}\)^{\!2}_{}\!e}{\,16\pi^2_{}M_\chi^{}\,}
\,.
\end{eqnarray}
Thus, we can estimate the decay width
\begin{eqnarray}
&&\Gamma(\chi_2^{} \!\to\! \chi_1^{} \gamma)
\sim \frac{\alpha^3}{\,\sin^4_{}\!\theta_W^{}\,}
\frac{(\Delta M_\chi^{})^3_{} }{M_\chi^2}
\,\simeq\, 8.5 \!\times\! 10^{-22}_{}\textrm{GeV}\!
\(\!\frac{\Delta M_\chi^{}}{\,550\,\textrm{keV}\,}\!\)^{\!\!3}_{} \!\!
\(\!\frac{M_\chi^{}}{\,1.2\,\textrm{TeV}\,}\!\)^{\!\!-2}_{}
\nonumber\\
&&
\gg H(T)\!\left|_{T=T_{\textrm{BBN}}^{}}^{}\right.
\!\simeq\, 4.5\!\times\! 10^{-25}_{}\textrm{GeV}\!  \(\!\frac{T_{\textrm{BBN}}^{}}{\,1\,\textrm{MeV}\,}\!\)^{\!\!2}_{}\,.
\label{eq:H-BBN}
\end{eqnarray}
This shows that the typical decay rate of $\,\chi_2^{}\!\to\! \chi_1^{}\gamma\,$
can be much faster than the BBN epoch for the mass-difference
$\,\Delta M_\chi^{}\!\sim 550\,$keV or larger.
In Eq.\eqref{eq:H-BBN}, the Hubble constant $H(T)$ is given by
\begin{eqnarray}
\label{hubble}
H(T)\,=\,\left(\!\frac{8\pi^{3}_{}g_{\ast}^{}}{90}\!\right)^{\!\!\!\frac{1}{2}}_{}\!
\frac{T^{2}_{}}{\,M_{\textrm{Pl}}^{}\,}\,,
\end{eqnarray}
with $M_{\textrm{Pl}}^{}\simeq 1.22\!\times\! 10^{19}_{}\,\textrm{GeV}$
being the Planck mass and
$\,g_{\ast}^{}=10.75$\, denoting the
relativistic degrees of freedom during the BBN epoch.
\vspace*{1mm}

The lighter Majorana fermion $\chi_1^{}$ will remain stable
and leave a relic density in the present universe.
Eqs.\eqref{mmsplit} and \eqref{cmsplit} show that the charged fermions $\tilde\xi^{\pm}_{}$
and neutral fermions $\chi_{1,2}^{}$ can be fairly quasi-degenerate.
Thus, for computing the relic density, we should take into account
not only the annihilations of the lightest $\chi_1^{}$,\,
but also the annihilations and co-annihilations involving the heavier
$\tilde\xi^\pm_{}$ and $\chi_{2}^{}$\,.\,
Such annihilation and co-annihilation processes can be induced by either gauge interactions
or Yukawa interactions.
As we will explain after Eq.\eqref{eq:sigmav}, our model can realize the case where
the gauge interactions dominate the annihilation and co-annihilation processes,
while the processes from Yukawa interactions can be negligible.
The processes induced by gauge interactions can contain in their final states
the electroweak gauge bosons, the Higgs bosons, and the fermions.
Thus, the corresponding effective cross section is a sum
of the leading contributions with the final states of gauge bosons (G),
Higgs bosons (H), and fermions (F),
%
\beqa
\label{eq:sigmav}
\left<\sigma v\right> \,\simeq\,
\left<\sigma v\right>_{\text{G}}^{}\! + \left<\sigma v\right>_{\text{H}}^{}
+\left<\sigma v\right>_{\text{F}}^{}
\,\simeq\,
\frac{~87\alpha_2^2+\,24\alpha_2^{}{\alpha}_Y^{}+\,45{\alpha}_Y^2~}
{64\pi^{-1}\!M_{\chi_1^{}}^2}\,,
\eeqa
%
where $\,(\alpha_2^{},\,\alpha_Y^{})=\frac{1}{4\pi}(g^2,\, g'^2)$,\,
and $(g,\, g')$ denote the weak and hypercharge gauge couplings,
respectively.  As we have checked, the above gauge interaction contributions
agree with \cite{cfs2006}.
The Yukawa interactions can induce the $s$-channel annihilations
$\,\chi_1^{} \chi_1^{} \!\!\to\! \phi\phi\,$ and
the $t$-channel annihilations
$\,\chi_1^{} \chi_1^{} \!\!\to\! \Delta \Delta^{\dag}$\,,\,
as well as the related co-annihilation channels.
In our analysis, we consider the Higgs triplet mass
to be around the scale of DM mass (with
$M_{\Delta}^{}< 2M_{\chi_1}^{}$),\,
and the Yukawa couplings to be reasonably small
[\,$f,f'= {\cal O}(0.1\!-0.01)$\,].\,
Thus, we find that the (co)annihilation processes
$\,\chi_{1,2}^{} \chi_{1,2}^{} \!\!\to\! \phi\phi\,$ are suppressed
by the product of the $\Delta\phi\phi$ coupling and $s$-channel propagator
factor, which is proportional to
$(\mu_{\Delta\phi}/M_\Delta^2)^2\propto v_{\Delta}^2/v_\phi^4$,\,
where $\,v_{\Delta}^2/v_\phi^2\ll 1\,$ [due to Eq.\eqref{htvev}] and
$M_{\Delta}^{}$ is chosen significantly away from the resonant production
of $\Delta$\,.\,
For the (co)annihilation processes such as
$\,\chi_1^{} \chi_1^{} \!\to \Delta \Delta^\dag$\, and
$\,\chi_1^{} \Delta \!\to \!\chi_1^{} \Delta$\,,\,
the effective cross sections will be suppressed
by the Yukawa coupling factors $(f^4,\,f^{'4},\,f^2f^{'2})$\, for
$f,f'\!\! = {\cal O}(0.1\!-0.01)$.\,
With the above consideration, our model realizes the conventional minimal DM scenario,
with gauge interactions dominating the DM annihilation and co-annihilation processes
as in Eq.\eqref{eq:sigmav}.

\vspace*{1mm}

For the current study, the corresponding DM relic density can be expressed as
\cite{kt1990,DMrelic},
\beqa
\label{eq:nDM/s}
\frac{n_{\chi_1^{}}^{}}{s} \,\simeq\,
\frac{~x_f^{}~}{\,\left<\sigma v\right>\!M_{\chi_1^{}}^{}\!M_{\text{Pl}}^{}\,}\!
\(\!\frac{\,180\,}{\,\pi \,g_*^{}\,}\)^{\!\!\frac{1}{2}},\,
\hspace*{10mm}
x_f^{}\,\simeq\,
\ln\frac{\,\left<\sigma v\right>\!M_{\chi_1^{}}^{}\!M_{\text{Pl}}^{}\,}{60\sqrt{g_*^{}}}\,,
\eeqa
where $\,n_{\chi_1^{}}^{}$ is the DM number density at the freeze-out temperature
$T_f^{}$, \,$s$\, is their total entropy,
$M_{\textrm{Pl}}^{}\simeq 1.22\times 10^{19}_{}\,\textrm{GeV}$
is the Planck mass, and the ratio
$\,x_f^{}={M_{\chi_1^{}}^{}}\!/{T_f^{}}\,$.\,
The quantity $g_*^{}$ denotes the effective relativistic degrees of freedom
in the thermal equilibrium at the freeze-out temperature, and
thus $\,g_*^{} = 106.75$\,.\,
For the $\Lambda$CDM cosmology, the latest Planck data give,
$\,\Omega_{\text{DM}}^{}h^2=0.120\pm 0.001$ \cite{Planck2018}.
With the generic relation,
\beqa
\frac{n_{\chi_1^{}}^{}}{s} \,=\,
\frac{\,0.436\,\text{eV}\,}{M_{\chi_1^{}}^{}}^{}
\frac{~\Omega_{\chi_1^{}}^{}\!h^2\,}{0.120}\,,
\eeqa
and using Eqs.\eqref{eq:sigmav}-\eqref{eq:nDM/s}, we compute
the mass of the lightest fermion $\chi_1^{}$ as
a stable DM particle,
\begin{eqnarray}
\label{dmmass}
M_{\chi_1^{}}^{}\simeq\, 1.24\,\textrm{TeV}\,.
\end{eqnarray}

\vspace*{1mm}

Due to the small VEV $\,|v_\Delta^{}|\leqq 3.1\,\textrm{GeV}$
and the value of rotation angle $\,\theta\simeq \fr{\pi}{4}$\,,\,
the spin-dependent elastic scattering of the DM particle $\chi_1^{}$
off the nucleon will be far below the experimental sensitivities.
As for the spin-independent elastic scattering,
its cross section can be computed at one-loop level
and estimated as\,\cite{cfs2006},
\begin{eqnarray}
\sigma_{\textrm{SI}}^{} \,\simeq\,
3\times\! 10^{-46}_{}\,\textrm{cm}^2_{}\,,
\end{eqnarray}
which is reachable by the future direct detection experiments \cite{DM-DirectExp-Rev}.
Furthermore, the DM particle $\chi_1^{}$ and its heavier partner $\chi_2^{}$
have a spin-independent inelastic scattering off the nucleon at tree-level.
If the $\,\chi_1^{}\!-\!\chi_2^{}\,$ mass-splitting would vanish,
the spin-independent inelastic scattering cross section
takes the following form\,\cite{jkg1996},
\begin{eqnarray}
\sigma_0^{}\!&=\!&
\frac{\,m_p^2 G_F^2\,}{128\pi}
\left[\frac{(A\!-\!Z)-(1\!-\!4\sin^2\!\theta_W)Z\,}{A}\right]^2_{}
\nonumber\\[1mm]
\!&=&\! 1.16\!\times\!\! 10^{-40}_{}\textrm{cm}^2_{}\!
\left[\frac{\,(A\!-\!Z)\!-\!(1\!-\!4\sin^2\!\theta_W)Z\,}{A}\right]^2_{}\!\!,~~~~~
\label{eq:sigma0}
\end{eqnarray}
and is already excluded by the direct DM searches.
In the above equation, $\,m_p^{}\simeq 938\,\textrm{MeV}$\, is the proton mass,
$G_F^{}\simeq 1.16638\times 10^{-5}_{}\,\textrm{GeV}^{-2}_{}$ is the Fermi constant,
and $(Z,\,A)$ denote the (charge,\,mass) numbers of the target nuclei, respectively.
On the other hand, our inelastic DM model predicts small
but nonzero $\,\chi_1^{}\!-\!\chi_2^{}\,$ mass-splitting
in the range of Eq.(\ref{mmsplit}) or Eq.(\ref{mmsplit2}).
Hence, the corresponding inelastic DM scattering
can escape from the current experimental constraints.
Other inelastic DM models have similar feature
regarding the experimental direct detection constraints,
as discussed in the literature\,\cite{tw2001,extra1-iDM,extra2-iDM,cktw2009}.

\vspace*{4mm}
\section{Radiative Type-II Neutrino Seesaw}
\vspace*{1.5mm}
\label{sec:4}

\begin{figure}[t]
\begin{center}
\includegraphics[width=6.5cm,height=5cm]{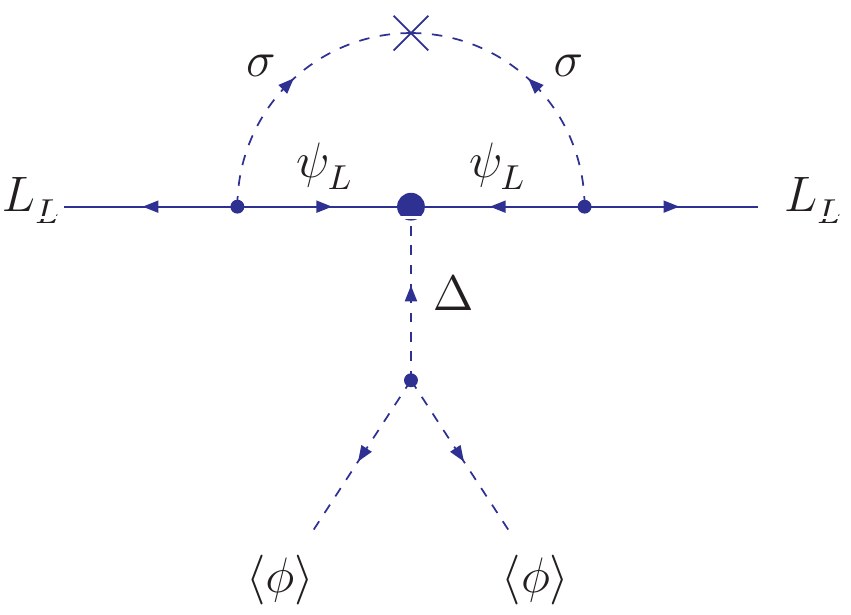}
\vspace*{-1mm}
\caption{One-loop diagram for generating the radiative neutrino masses
through type-II seesaw in the present model.}
\vspace*{-2mm}
\label{fig:1}
\end{center}
\end{figure}

After the electroweak symmetry breaking, we can generate
a Majorana mass term for the left-handed neutrinos
via radiative type-II seesaw, as given by the one-loop diagram in Fig.\,\ref{fig:1},
\begin{eqnarray}
\mathcal{L}_{\text{mass}}^{\nu} \,=\, -\fr{1}{2}\overline{\nu_L^c}
\mathbf{m}_\nu^{}\nu_L^{}\!+\textrm{H.c.}
\,=\, \fr{1}{2}\nu^T \mathbf{m}_\nu^{}\nu +\textrm{H.c.},~~
\end{eqnarray}
where $\,\nu_L^{}=\(\!\ba{c} \nu \\ 0\ea\!\)\,$
and $\,\nu\,$ is a left-handed
Weyl spinor in the $(\frac{1}{2},\,0)$ representation of Lorentz group.
We note that in Fig.\,\ref{fig:1}, the lepton number violation (\,$\Delta L=2$\,)
arises from the $\sigma$-propagator with the mass-insertion
$\,\tilde{M}^2_{\sigma j}\propto M^2_{\sigma jR} - M^2_{\sigma jI}\,$.\,
The Majorana nature of the neutrino mass-generation originates from the vertices
$\Delta\!-\psi_L^{}\!-\psi_L^{}$ and $\Delta^{\dag}\!-\psi_L'\!-\psi_L'$.\,
From Fig.\,\ref{fig:1},
we compute the one-loop radiative neutrino masses as follows,
\begin{eqnarray}
\label{numassform}
(\mathbf{m}_\nu^{})_{\alpha\beta}^{}
&\!\!=\!\!&
\frac{M_{\chi_1^{}}^{}}{\,16\pi^2_{}\,}
\sum_{j}^{}y_{\alpha j}^{}y_{\beta j}^{}\!\!
\left[\frac{M_{\sigma^{}_{jR}}^2}{\,M_{\sigma^{}_{jR}}^2\!\!-\!M_{\chi_1^{}}^2}
\!\ln\!\(\!\!\frac{\,M_{\sigma^{}_{jR}}^2}{M_{\chi_1^{}}^2}\!\!\)
-\frac{M_{\sigma^{}_{jI}}^2}{\,M_{\sigma^{}_{jI}}^2\!\!-\!M_{\chi_1^{}}^2}
\ln\!\left(\!\!\frac{\,M_{\sigma^{}_{jI}}^2}{M_{\chi_1^{}}^2}\!\!\)\!\right]
\nonumber\\[1mm]
&&
-\frac{M_{\chi_2^{}}^{}}{16\pi^2_{}}\sum_{j}^{}y_{\alpha j}^{}
y_{\beta j}^{}\!\!\left[\frac{M_{\sigma^{}_{jR}}^2}
{\,M_{\sigma^{}_{jR}}^2\!\!-M_{\chi_2^{}}^2\,}
\!\ln\!\(\!\!\frac{\,M_{\sigma^{}_{jR}}^2}{M_{\chi_2^{}}^2}\!\!\)
-\frac{M_{\sigma^{}_{iI}}^2}{\,M_{\sigma^{}_{jI}}^2\!\!-M_{\chi_2^{}}^2\,}
\!\ln\!\(\!\!\frac{\,M_{\sigma^{}_{jI}}^2}{M_{\chi_2^{}}^2}\!\!\)\!\right]\!.
\end{eqnarray}

As we will show later, we are interested in the case
where the scalars $\,(\sigma_{jR}^{},\,\sigma_{jI}^{})$\, can realize the high-scale
leptogensis and thus are much heavier than the fermions $(\chi^{}_{1},\,\chi_2^{})$.\,
In this case, we can simplify the above mass formula as
\begin{eqnarray}
\label{numassform2}
(\mathbf{m}_\nu^{})_{\alpha\beta}^{}
\,\simeq\, \frac{\,\Delta M_{\chi}^{}\,}{16\pi^2_{}}
\sum_{j}^{}y_{\alpha j}^{}y_{\beta j}^{}
\!\ln\!\(\!\!\frac{\,M_{\sigma^{}_{jR}}^2\,}{M_{\sigma_{jI}^{}}^2}\!\!\)\!,
~~~~~
\end{eqnarray}
for $\,M_{\sigma_{jR}^{}}^2\!\neq M_{\sigma_{jI}^{}}^2 \!\gg M_{\chi_{1,2}^{}}^2$.\,
Inspecting Eqs.\eqref{numassform} and \eqref{numassform2}, we see that the neutrino mass
$\mathbf{m}_\nu^{}$\! vanishes {\it if} $\,M_{\sigma_{jR}^{}}^{}\!\!= M_{\sigma_{jI}^{}}^{}$.\,
This is expected because the lepton number violation ($\Delta L=2$)
of the neutrino mass term
is generated by the mass-insertion of the $\sigma$ field in Fig.\,\ref{fig:1},
$\,\tilde{M}^2_{\sigma j}\propto M^2_{\sigma jR} - M^2_{\sigma jI}\,$.\,
Furthermore, Eq.\eqref{numassform} shows that $\mathbf{m}_\nu^{}\!=0$\,
if $\,M_{\chi_1^{}}^{}\!\!= M_{\chi_2^{}}^{}$,\, since the contributions of
$\chi_1^{}$ and $\chi_2^{}$ to the loop diagram of Fig.\,\ref{fig:1}
take the same form but with opposite signs.
This is manifest in Eq.\eqref{numassform2}
where the neutrino mass is proportional to the mass-difference
$\,\Delta M_{\chi}^{}\,$ and thus the triplet VEV,
$\,\mathbf{m}_\nu^{}\!\propto \Delta M_{\chi}^{}\!\propto v_\Delta^{}\,$.\,
In addition, we note that for $\,M_{\sigma_{jR,I}^{}}^2\!\! \gg M_{\chi_{1,2}^{}}^2$,\,
the mass-scale of the loop diagram in Fig.\,\ref{fig:1}
is controlled by the heavy mass of $\sigma_j^{}$,\,
so the resultant approximate neutrino-mass formula \eqref{numassform2}
is controlled by the heavy $\sigma_j^{}$ mass and does not depend on
the small TeV-scale masses of $(\chi^{}_{1},\,\chi_2^{})$,\,
except the overall mass-difference factor $\,\Delta M_{\chi}^{}\!\propto v_\Delta^{}\,$
due to the VEV of the external triplet field $\Delta$\,.
This feature is important for our following interpretation of neutrino mass-generation via
type-II seesaw around Eqs.\eqref{eq:Lseesaw-II}-\eqref{eq:feff-seesaw-II}.

To generate the required size of neutrino masses
$\,m_\nu^{}=\mathcal{O}(0.1\textrm{eV})$, we set the Yukawa couplings as
\,$y\gtrsim \mathcal{O}(10^{-4}_{})$\,
for $\,\Delta M_\chi^{}\lesssim 17.1\,\textrm{GeV}$,\,
or, $\,y\gtrsim \mathcal{O}(10^{-2}_{})$\,
for $\,\Delta M_\chi^{}\lesssim 550\,\textrm{keV}$,\,
since the logarithm function usually has a value of $\mathcal{O}(1)$.
Note that here we have a rank-2 neutrino mass matrix
$\,\mathbf{m}_\nu^{}$ with two nonzero eigenvalues
because the model only contains two scalar singlets $\sigma_{1,2}^{}$
and their Yukawa couplings $\,y_{\alpha j}^{}$ form a $3\!\times\!2$ matrix.
(This feature is similar to the minimal type-I neutrino seesaw
with two right-handed heavy Majorana neutrinos\,\cite{MSS-I}.)
If the light neutrinos have three nonzero mass eigenvalues,
then we can readily extend the present model with three singlet scalars
$\sigma_j^{}$ ($j=1,2,3$), which will not affect the main feature of the
present model.

\vspace*{1mm}

The neutrino mass-generation (\ref{numassform2}) may be also understood
as a type-II seesaw. Since the scalar singlets $(\sigma_1^{},\,\sigma_2^{})$
have masses around the leptogenesis scale (cf.\ Sec.\,\ref{sec:5}) and
are extremely heavy, we can expand the exact mass formula \eqref{numassform}
to obtain Eq.\eqref{numassform2}.
As we explained below Eq.\eqref{numassform2}, the loop diagram of Fig.\,\ref{fig:1}
is controlled by the heavy mass of the singlet $\sigma_j^{}$ and the resultant approximate
formula \eqref{numassform2} does not depend on the small masses of the light fields
$(\chi_1^{},\,\chi_2^{})$ in the loop diagram, except the overall mass-difference factor
$\,\Delta M_{\chi}^{}\!\propto v_\Delta^{}\,$
due to the VEV of the external triplet field $\Delta$\,.\,
Thus, it is instructive to view this as integrating out the
heavy singlets $(\sigma_1^{},\,\sigma_2^{})$, and we obtain
the low energy effective Yukawa interactions
between the Higgs triplet $\Delta$ and the SM lepton doublets $L_L^{}$,
\beqa
\label{numassform3}
\label{eq:Lseesaw-II}
\mathcal{L}^Y_{\text{eff}} ~=~
-\frac{1}{2}f_{\textrm{eff}}^{}\,\overline{L_{L}^c} \ii\tau_2^{}\Delta L_L^{}
+\textrm{H.c.},
\eeqa
where $f_{\textrm{eff}}^{}$ is the effective Yukawa coupling with
\beqa
\label{eq:feff-seesaw-II}
(f_{\textrm{eff}}^{})_{\alpha\beta}^{}
~=~ \frac{\,f\!+\!f'\,}{16\pi^2_{}}\sum_{j}^{}
y_{\alpha j}^{}y_{\beta j}^{}
\ln\!\(\!\!\frac{\,M_{\sigma^{}_{jR}}^2}{\,M_{\sigma_{jI}^{}}^2}\!\!\),
~~~~~
\eeqa
for $\,M_\Delta^{}\ll M_{\sigma^{}_{jR}}^{}\!\neq M_{\sigma_{jI}^{}}^{}$.\,
We may call the above as a radiative type-II neutrino seesaw
since it is realized at one-loop level.
For such neutrino mass-generation with a fairly small Higgs triplet VEV
$\,v_\Delta^{}\lesssim 0.1\,\textrm{MeV}$,\,
it can be tested at the LHC \cite{fhhlw2008}.

As a further remark, we inspect the effective coupling \eqref{eq:feff-seesaw-II} and
note that $\,f_{\text{eff}}^{}$\,  and the corresponding Majorana neutrino
mass would vanish if the mass-splitting
$\,M_{\sigma^{}_{jR}}^2\!\!- M_{\sigma_{jI}^{}}^2\!
 (=2\tilde{M}^2_{\sigma j})$\, equals zero.
This is because the mass-splitting between the real and imaginary components
of $\sigma_j^{}$  arises from the soft lepton number breaking
via the mass-term of $\tilde{M}^2_{\sigma j}$,\,
as we explained below Eq.\eqref{numassform2}.
When the $\sigma_j^{}$ mass goes to infinity for any
fixed finite mass-difference between $(\sigma_{jR}^{},\sigma_{jI}^{ })$,
we see that the effective coupling \eqref{eq:feff-seesaw-II} also approaches zero,
which is consistent with the decoupling theorem.

\vspace*{4mm}
\section{Realizing Natural High-Scale Leptogenesis}
\vspace*{1.5mm}
\label{sec:5}

\begin{center}
\begin{figure*}[t]
\includegraphics[width=17cm,height=6.5cm]{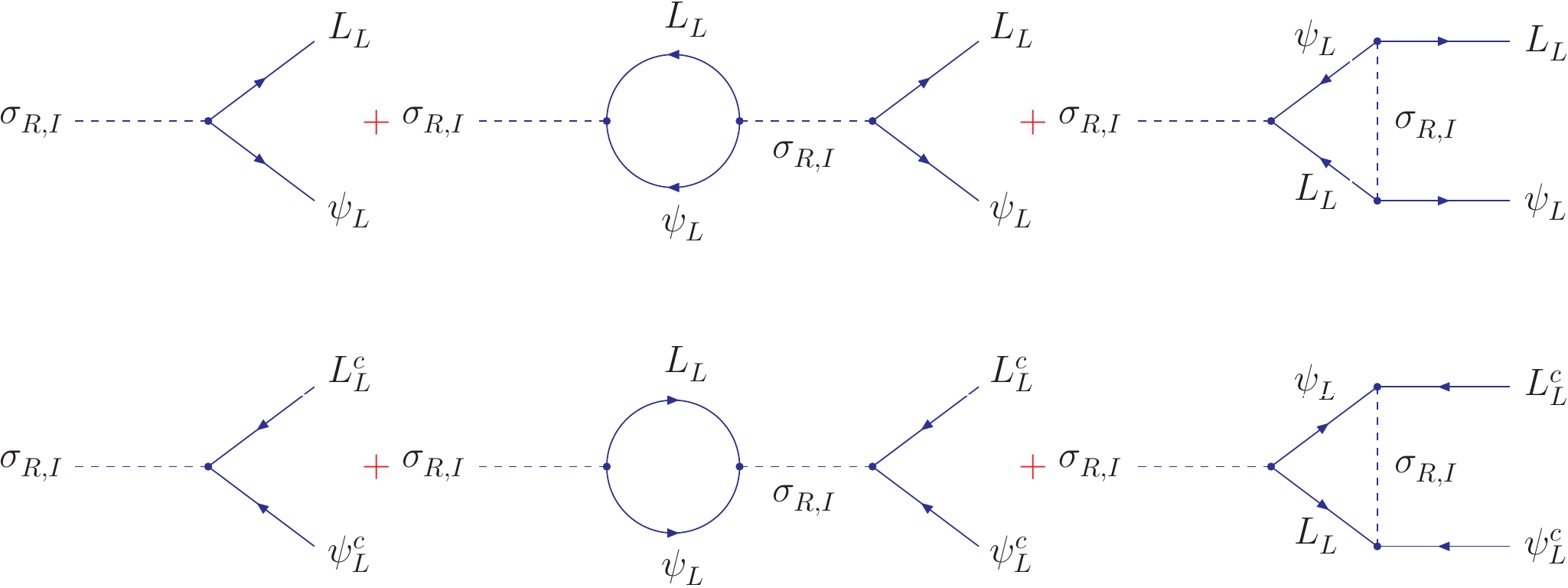}
\caption{Decays of the singlet scalars
$\,\sigma_j^{}=\frac{1}{\sqrt{2}}(\sigma_{R}^{}+\ii\sigma_{I}^{})$\,
into the SM lepton doublets $L_L^{}$ and the new fermion doublet $\psi_L^{}$.}
\label{fig:2}
\vspace*{2mm}
\end{figure*}
\end{center}

In the present model, we can realize a leptogenesis through the decays of the real scalars
$\sigma_{iR}^{}$ or $\sigma_{iI}$.
The relevant Feynman diagrams are shown in Fig.\,\ref{fig:2}.
Thus, we compute the decay widths at tree level,
\beqs
\begin{eqnarray}
\Gamma_{\sigma_{iR}^{}}^{} &\!\!\equiv\!\!&
\sum_{\alpha }[\Gamma(\sigma_{iR}^{}\!\rightarrow\!
L_{L\alpha}^{}\!\!+\!\psi_{L}^{})+
\Gamma(\sigma_{iR}^{}\!\rightarrow\! L_{L\alpha}^{c}\!\!+\!\psi_{L}^{c})]
\,=\, \frac{1}{\,8\pi\,}(y^\dagger_{}y)_{ii}^{}M_{\sigma_{iR}^{}}^{},
\\[3mm]
\Gamma_{\sigma_{iI}^{}}^{} &\!\!\equiv\!\!&
\sum_{\alpha }[\Gamma(\sigma_{iI}^{}\!\rightarrow\! L_{L\alpha}^{}\!\!+\!\psi_{L}^{})
+\Gamma(\sigma_{iI}^{}\!\rightarrow\! L_{L\alpha}^{c}\!\!+\!\psi_{L}^{c})]
\,=\, \frac{1}{\,8\pi\,}(y^\dagger_{}y)_{ii}^{}M_{\sigma_{iI}^{}}^{}.
\end{eqnarray}
\eeqs
Then, we evaluate the CP asymmetries at one-loop order,
\beqs
\begin{eqnarray}
\varepsilon_{\sigma_{iR}^{}}^{}
&\!\!\equiv\!\!&
\frac{\,\sum_{\alpha }[\Gamma(\sigma_{iR}^{}\!\rightarrow\! L_{L\alpha}^{}\!\!+\!\psi_{L}^{})
-\Gamma(\sigma_{iR}^{}\!\rightarrow\! L_{L\alpha}^{c}\!\!+\!\psi_{L}^{c})]\,}
{\Gamma_{\sigma_{iR}^{}}^{}}
\nonumber\\[1mm]
&\!\!=\!\!&
\frac{1}{24\pi}\frac{\textrm{Im}\{[(y^\dagger_{}y)_{ij}^{}]^2_{}\}}{(y^\dagger_{}y)_{ii}^{}}
\left\{\!\left[S\!\left(\!\frac{M_{\sigma_{jR}^{}}^2}{M_{\sigma_{iR}^{}}^2}\!\!\right)\!
+V\!\left(\!\frac{M_{\sigma_{jR}^{}}^2}{M_{\sigma_{iR}^{}}^2}\!\!\right)\right]
-\left[S\!\left(\!\frac{M_{\sigma_{jI}^{}}^2}{M_{\sigma_{iR}^{}}^2}\!\!\right)\!
+V\!\left(\!\frac{M_{\sigma_{jI}^{}}^2}{M_{\sigma_{iR}^{}}^2}\!\!\right)\right]\!\right\} ,
\\[3mm]
\varepsilon_{\sigma_{iI}^{}}^{} &\!\!\equiv\!\!&
\frac{\,\sum_{\alpha}[\Gamma(\sigma_{iI}^{}\!\!\rightarrow\! L_{L\alpha}^{}\!\!+\!\psi_{L}^{})
-\Gamma(\sigma_{iI}^{}\!\!\rightarrow L_{L\alpha}^{c}\!\!+\!\psi_{L}^{c})]\,}
{\Gamma_{\sigma_{iI}^{}}^{}}
\nonumber
\\[1mm]
&\!\!=\!\!&
-\frac{1}{24\pi}
\sum_{j\neq i}^{}\frac{\textrm{Im}\{[(y^\dagger_{}y)_{ij}^{}]^2_{}\}}{(y^\dagger_{}y)_{ii}^{}}
\left\{\!\left[S\!\left(\!\frac{M_{\sigma_{jR}^{}}^2}{M_{\sigma_{iI}^{}}^2}\!\!\right)\!
+V\!\left(\!\frac{M_{\sigma_{jR}^{}}^2}{M_{\sigma_{iI}^{}}^2}\!\!\right)\right]\!
-\left[S\!\left(\!\frac{M_{\sigma_{jI}^{}}^2}{M_{\sigma_{iI}^{}}^2}\!\!\right)\!
+V\!\left(\!\frac{M_{\sigma_{jI}^{}}^2}{M_{\sigma_{iI}^{}}^2}\!\!\right)\right]\!\right\}\! ,
\hspace*{13mm}
\end{eqnarray}
\eeqs
where $\,S(x)\,$ and $\,V(x)\,$ are the self-energy and vertex corrections, respectively,
\beqs
\begin{eqnarray}
S(x)&=&\frac{2}{\,x\!-\!1\,}\,,
\\
V(x)&=& (1\!+\!2x)\!\!\left[2+(1\!+\!2x)\ln\!\frac{x}{\,1\!+\!x\,}\right]\!.~~~~~~~~
\end{eqnarray}
\eeqs
%

As an example, we set the masses of the four real scalars
$(\sigma_{1R}^{},\,\sigma_{1I}^{})$ and $(\sigma_{2R}^{},\,\sigma_{2I}^{})$
with the following hierarchy,
\begin{eqnarray}
M_{\sigma_{1I}^{}}^2\ll\, M_{\sigma_{1R}^{}}^2
\ll\, M_{\sigma_{2I}^{}}^2
\ll\, M_{\sigma_{2R}^{}}^2\,.
\end{eqnarray}
Thus, the final baryon asymmetry should mainly come from the decays
of the lightest singlet scalar $\sigma_{1I}^{}$.\,
In this case, we find that the relevant CP asymmetry becomes
\begin{eqnarray}
\varepsilon_{\sigma_{1I}^{}}^{}
\simeq\, \frac{1}{\,8\pi\,}
\frac{\,\textrm{Im}\{[(y^\dagger_{}y)_{12}^{}]^2_{}\}\,}{(y^\dagger_{}y)_{11}^{}}
\frac{\,M_{\sigma_{1I}^{}}^2\,}{\,M_{\sigma_{2I}^{}}^2\,}\,.
\end{eqnarray}

Note that due to the trilinear scalar vertex
$\phi\phi\Delta$ in Eq.\eqref{eq:L}
and the radiative Yukawa vertex
$\,L_L^{}L_L^{}\Delta$\, in Eq.\eqref{numassform3},
the Higgs triplet $\,\Delta\,$ will mediate some $\,\Delta L =2\,$ processes:
\,$L_{L}^{}L_{L}^{}\longleftrightarrow \phi\phi$,\,
\,$L_{L}^{c}L_{L}^{c}\longleftrightarrow \phi^{\ast}_{} \phi^{\ast}_{}$,\,
and \,$L_{L}^{}\phi_{}^{\ast} \longleftrightarrow L_{L}^{c} \phi $\,.\,
Before the sphalerons (for leptogenesis) stop working,
these additional lepton-number-violating processes
should keep out of the equilibrium and thus do not wash out the produced lepton
asymmetry from $\sigma_{1I}^{}$ decays. This will require
\begin{eqnarray}
\label{con}
\Gamma_{\Delta L=2}^{}<  H(T) \,,
\end{eqnarray}
for $\,T>T_{\textrm{sph}}^{}$\,.\,
Here the Hubble constant $H(T)$ is given by
%
$\,H(T)\,=\left(\!\frac{8\pi^{3}_{}g_{\ast}^{}}{90}\!\right)^{\!\!\!\frac{1}{2}}_{}\!
\frac{T^{2}_{}}{\,M_{\textrm{Pl}}^{}\,}$\,,\,
%
with $M_{\textrm{Pl}}^{}\simeq 1.22\times$ $10^{19}_{}\,\textrm{GeV}$
being the Planck mass and
$\,g_{\ast}^{}=\mathcal{O}(100)$\, denoting the
relativistic degrees of freedom during the leptogenesis epoch.
As for the rate of lepton-number-violating interactions, $\,\Gamma_{\Delta L=2}^{}$\,,\,
we can estimate
\beqs
\begin{eqnarray}
\label{lphi1}
\Gamma_{\Delta L=2}^{} &\!\sim\!\!&
\frac{\textrm{Tr}(f^\dagger_{\textrm{eff}}f^{}_{\textrm{eff}})
|\mu_{\Delta \phi}^{}|^2_{}}{T}=
\frac{4 M_\Delta^4 \textrm{Tr}(m^\dagger_{\nu} m^{}_{\nu})}{v_\phi^4\,T}\,,
~~~~~~(\textrm{for}~T>M_{\Delta}^{}),~~~~~~~
\\[2mm]
\label{lphi2}
\Gamma_{\Delta L=2}^{} &\!\sim\!\!&
\frac{\textrm{Tr}(f^\dagger_{\textrm{eff}}f^{}_{\textrm{eff}})
|\mu_{\Delta\phi}^{}|^2_{}T^3_{}}{M_{\Delta}^{4}}=
\frac{4\textrm{Tr}(m^\dagger_{\nu} m^{}_{\nu})T^3_{}}{v_\phi^4 }\,,
~~~ \hspace*{3.5mm}
(\textrm{for}~T<M_{\Delta}^{}).
\end{eqnarray}
\eeqs
%

\vspace*{-5mm}
For a numerical demonstration, we define
\begin{eqnarray}
\label{rwidth}
K&=&
\left.\frac{\Gamma_{\sigma_{1I}^{}}^{}}{\,2H(T)\,}\right|_{T=M_{\sigma_{1I}^{}}^{}}^{} \,,
\end{eqnarray}
where $H(T)$ is the Hubble constant given by Eq.\eqref{hubble}.
In the weak washout region, the final baryon asymmetry
can be described as follows\,\cite{kt1990},
\begin{eqnarray}
\eta_B^{}~=~ \frac{\,n_B^{}}{s}
\,\simeq\,
-\frac{28}{79}\!\times\! \frac{\,\varepsilon_{\sigma_{1I}^{}}^{}}{g_\ast^{}} \,,
\end{eqnarray}
for $\,K\ll 1$\,.\,
Here $n_B^{}$ and $s$ denote the baryon number density and the entropy density, respectively,
while the factor $-\frac{28}{79}\,$ is the sphaleron lepton-to-baryon coefficient.
In the present model, we have $\,g_\ast^{}=119.75$\,
accounting for the SM fields plus one Higgs triplet $(\Delta)$
and two fermion doublets $(\psi_L^{}$ and $\psi_L')$.
The latest Planck observation gives\,\cite{Planck2018},
\beqa
\label{eq:etaB-data}
\eta_B^{} \,=\, (6.12\pm 0.03)\!\times\! 10^{-10}\,.
\eeqa

\begin{figure}[t]
\begin{center}
\hspace*{-5mm}
\includegraphics[width=10.5cm,height=7cm]{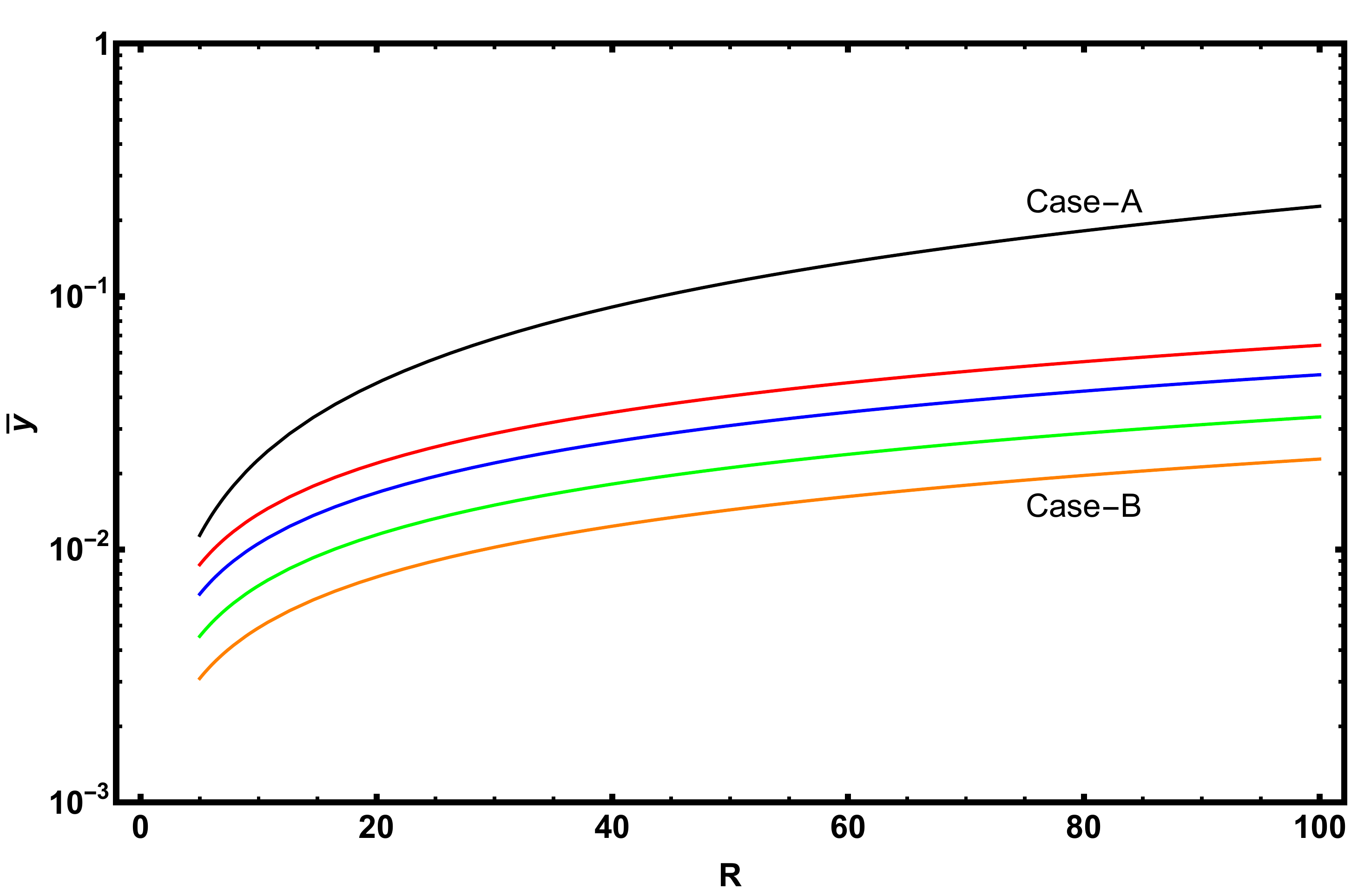}
\vspace*{-2mm}
\caption{The characteristic size of effective Yukawa coupling $\,\overline{y}\,$ as
a function of the scalar mass ratio
$R=M_{\sigma_{2I}^{}}^{}\!/M_{\sigma_{1I}^{}}^{}$ under
the constraint of baryon asymmetry \eqref{eq:etaB-data},
where $\,\overline{y}=\bar{y}_{\text{TL}}^{}\,$ is defined in Eq.\eqref{eq:yb-A} for Case-A
and $\,\overline{y}=\bar{y}_{\text{nTL}}^{}\,$ in Eq.\eqref{eq:yb-B} for Case-B.
In the Case-B, the (yellow, green, blue, red) curves
from the bottom to top, correspond to the lowest singlet scalar mass
$\,M_{\sigma_{1I}^{}}^{}\!\!=(10^{13},\,10^{14},\,10^{15},\,5\!\times\!10^{15})$GeV.
}
\label{fig:3}
\vspace*{-7mm}
\end{center}
\end{figure}

For illustration, it is useful to define a simple effective coupling
$\,\yb_{\text{TL}}^{}\,$ to characterize the {\it size} of the relevant Yukawa couplings
for the $L_L^{}$-$\psi_L^{}$-$\sigma$ vertex (especially, its order of
magnitude), without invoking the detailed structure of Yukawa matrix,
\beqa
\label{eq:yb-A}
\yb^2_{\text{TL}} \equiv
-\frac{\,\textrm{Im}\{[(y^\dagger_{}y)_{12}^{}]^2_{}\}\,}
  {(y^\dagger_{}y)_{11}^{}} \,.
\eeqa
Thus, we have
\beqa
\yb_{\text{TL}}^{} ~\simeq\,
\sqrt{\frac{\,158g_*^{}\eta_B^{}\,}{7}\,}\!\!
\(\!\!\frac{\,M_{\sigma_{2I}^{}}^{}}{\,M_{\sigma_{1I}^{}}^{}}\!\!\).
\eeqa
With this, we impose the baryon asymmetry data \eqref{eq:etaB-data}
and estimate the allowed range of the effective Yukawa coupling
$\,\yb_{\text{TL}}^{}$ as a function of the scalar mass-ratio
$\,R=M_{\sigma_{2I}^{}}^{}\!/M_{\sigma_{1I}^{}}^{}$.\,
We plot this as the black curve (called Case-A)
in Fig.\,\ref{fig:3} for the range of
$\,5\leqq R\leqq 10^2\,$.
We input the experimental central value of Eq.\eqref{eq:etaB-data}.
We also vary the value of $\,\eta_B^{}$ within $\pm 3\sigma$
range, but find no visible effect in Fig.\,\ref{fig:3}.
From Fig.\,\ref{fig:3}, we see that the typical size of
Yukawa couplings (\,$\bar{y}_{\text{TL}}^{}$\,) can be naturally around
${\cal O}(10^{-1}\!-\!10^{-2})$.

For an explicit numerical sample, we can choose the sample inputs,
\begin{eqnarray}
\label{pchoice}
&&M_{\sigma_{1I}^{}}^{}=10^{-1}M_{\sigma_{2I}^{}}^{}
\!=1.5\!\times\! 10^{13}_{}\textrm{GeV},
\nonumber\\[1mm]
&&M_{\sigma_{1R}^{}}^{}=10^{-1}M_{\sigma_{2R}^{}}^{}
\!=7.5\!\times\! 10^{13}_{}\textrm{GeV},~~~~~~
\nonumber\\[1mm]
&&\Delta M_\chi^{}=550\,\textrm{keV},~~
\yb_{\text{TL}}^{} =2.27\!\times\! 10^{-2}_{}\,.
\end{eqnarray}
Thus, we have
\begin{eqnarray}
\varepsilon_{\sigma_{1I}^{}}^{} \!\simeq\, 2.1\!\times\! 10^{-7},
~~~~~~~
\eta_B^{} \,\simeq\, 6.1\!\times\! 10^{-10}\,,
\end{eqnarray}
where the produced baryon asymmetry $\,\eta_B^{}\,$ is
consistent with the recent Planck observation\,\cite{Planck2018}
in Eq.\eqref{eq:etaB-data}.
We can also estimate $K=\mathcal{O}(0.1)$.\,

\vspace*{1.5mm}

As another approach to realize a non-thermal leptogenesis,
we may use the imaginary scalar component $\sigma_{1I}^{}$
to play the role of an inflaton \cite{klw2014}.
In this case, a final baryon asymmetry can be induced\,\cite{kt1990},
\begin{eqnarray}
\eta_B^{}\,=\,
-\frac{28}{79}\frac{T_{\textrm{RH}}^{}}{\,M_{\sigma_{1I}}^{}\,}\,
\varepsilon_{\sigma_{1I}^{}}^{} \,,
\end{eqnarray}
with $T_{\textrm{RH}}^{}$ being the reheating temperature \cite{kt1990},
\begin{eqnarray}
T_{\textrm{RH}}^{}&\equiv &T (t=\Gamma_{\sigma_{1I}^{}}^{-1})
\nonumber\\
&=&\left(\!\frac{90}{\,8\pi^{3}_{}g_{\ast}^{}\,}\!\!\right)^{\!\!\frac{1}{4}}_{}\!\!
\sqrt{\,\frac{(y^\dagger_{}y)_{11}^{}M_{\textrm{Pl}}^{}M_{\sigma_{1I}^{}}^{}\,}{16\pi}}\,.
\end{eqnarray}
For illustration of this model,
we may define a simple effective coupling
$\,\yb_{\text{nTL}}^{}\,$ to characterize the {\it size} of the relevant Yukawa couplings
for the $L_L^{}$-$\psi_L^{}$-$\sigma$ vertex,
\beqa
\label{eq:yb-B}
\yb^3_{\text{nTL}} \,\equiv\, -\frac{\,\textrm{Im}\{[(y^\dagger_{}y)_{12}^{}]^2_{}\}\,}
  {\sqrt{(y^\dagger_{}y)_{11}}\,} \,.
\eeqa
Thus, we can express $\,\yb_{\text{nTL}}^{}\,$ as follows,
\beqa
\yb_{\text{nTL}}^{} \,=
\left[\eta_B^{}\frac{158\pi}{7}\!
\(\!\!\frac{\,8\pi^3g_*^{}\,}{90}\!\!\)^{\!\!\!\frac{1}{4}}\!\!\!
\(\!\frac{16\pi M_{\sigma_{1I}^{}}^{}}{M^{}_{\text{Pl}}}\!\)^{\!\!\!\frac{1}{2}}
\right]^{\!\!\frac{1}{3}}\!\!\!
\(\!\frac{M_{\sigma_{2I}^{}}^{}}{M_{\sigma_{1I}^{}}^{}}\!\)^{\!\!\!\frac{2}{3}}\!\!.
\hspace*{5mm}
\eeqa
Next, imposing the baryon asymmetry data \eqref{eq:etaB-data},
we can estimate the allowed range of the effective Yukawa coupling
$\,\yb_{\text{nTL}}^{}\,$ as a function of the scalar mass-ratio
$\,R=M_{\sigma_{2I}^{}}^{}\!/M_{\sigma_{1I}^{}}^{}$.\,
In Fig.\,\ref{fig:3}, we plot this as the lower set of curves
in (yellow, green, blue, red) colors
(called Case-B) from the bottom to top, corresponding to
$\,M_{\sigma_{1I}^{}}^{}=(10^{13},\,10^{14},\,10^{15},\,5\!\times\!10^{15})$GeV,
for $\,5\leqq R\leqq 10^2\,$.

\vspace*{1mm}

For an explicit numerical sample of this model, we make
the following parameter choice,
\begin{eqnarray}
\label{pchoice2}
&& M_{\sigma_{1I}^{}}^{}\! =\, 20^{-1}\!M_{\sigma_{2I}^{}}^{}\!
=\, 4.5\!\times\!10^{13}_{}\textrm{GeV},
\nonumber\\
&& M_{\sigma_{1R}^{}}^{}\! =\, 20^{-1}\!M_{\sigma_{2R}^{}}^{}\!
=\, 2.3\!\times\!10^{14}_{}\textrm{GeV},~~~~~
\nonumber\\
&& \Delta M_\chi^{}=550\,\textrm{keV},~~~ \yb_{\text{nTL}}^{} =10^{-2}_{},
\end{eqnarray}
and derive
\beqa
\eta_B^{} \,\simeq\, 6.1\!\times\! 10^{-10}\,,
\eeqa
which is consistent with the Planck observation \eqref{eq:etaB-data} \cite{Planck2018}.
For illustration, we further choose a typical input
$(y^\dag y)_{11}^{1/2}\!=10^{-2}$\, and estimate
\begin{eqnarray}
\varepsilon_{\sigma_{1I}^{}}^{} \!\simeq\, 10^{-9}\,,~~~~~~
T_{\textrm{RH}}^{}\simeq\, 7.6\!\times\!10^{12}_{}\,\textrm{GeV}\,.
\end{eqnarray}

In the above samples, we choose $\Delta M_\chi^{}=550$\,keV and
the effective Yukawa coupling
$\,\yb_{\text{TL}}^{}, \yb_{\text{nTL}}^{} =\mathcal{O}(10^{-2})$.
Thus, from the radiative mass formula
\eqref{numassform2} of light neutrinos, we see that the
light neutrino mass scale of $\,m_\nu^{}=\mathcal{O}(0.1\text{eV})\,$
can be realized, which is consistent with the oscillation data\,\cite{olive2014}
and cosmological constraints\,\cite{Planck2018}.
Also, for the above numerical samples,
we have checked that the condition ({\ref{con})
is satisfied for the parameter choice (\ref{pchoice}) and (\ref{pchoice2}).
In particular, Eq.(\ref{lphi1}) will match the condition (\ref{con})
during the leptogenesis epoch $\,T\gg M_\Delta^{}\,$ and
the low temperature period $\,T\!\sim\! M_\Delta^{}$.\,
When the temperature falls to $\,T< M_\Delta^{}$,\,
the rate will become as Eq.(\ref{lphi2}) and
is reduced by a factor $\,T^4_{}/M_\Delta^{4}<1$.\,
This means that Eq.(\ref{lphi2}) also matches the condition (\ref{con}).

\vspace*{4mm}
\section{Conclusions}
\vspace*{1.5mm}
\label{sec:6}

Understanding the origins of the neutrino masses, the baryon asymmetry,
and the dark matter
altogether poses an important challenge to the particle physics today.
In the conventional seesaw framework, the neutrino mass generation and the leptogenesis for
baryon asymmetry are tied to the same high energy scale.
This means that a low-scale neutrino mass generation
could not be consistent with a high-scale leptogenesis.
In the present work, we demonstrated an attractive new possibility
that a radiative neutrino mass generation can be achieved at the TeV scale,
while a thermal or inflationary leptogenesis naturally happens at
the high scale. Furthermore, our model realizes a viable minimal inelastic
dark matter (DM) at the TeV scale, where the
mass-splitting between the DM particle and its heavier partner can be naturally generated
by the interactions related to the neutrino mass generation.

\vspace*{1mm}

In section\,\ref{sec:2}, we presented the model construction, which
extends the standard model with two gauge-singlet scalars
$(\sigma_1^{},\,\sigma_2^{})$,\,
a vector-like iso-doublet fermion $(\psi_L^{},\,{\psi_L'}^c)$,
and one iso-triplet Higgs $\Delta$\,. This model holds a softly broken
lepton number and an exactly conserved $\ZZ_2^{}$ discrete symmetry.
Then, in section\,\ref{sec:3}, we demonstrated that the lighter Majorana fermion $\chi_1^{}$
can serve as a stable DM candidate and provide the observed relic density in the present
universe with its mass $M_{\chi_1^{}}^{}\!\!\simeq 1.24$\,TeV.\,
This fermionic DM $\chi_1^{}$ can be searched by the current
direct/indirect DM detection experiments\,\cite{DM-DirectExp-Rev}
and by the on-going LHC experiments as well as the future high energy
$pp$ colliders\,\cite{pp}.
In Section\,\ref{sec:4}, we further demonstrated how our model can naturally realize
the minimal type-II seesaw and radiatively generate the light neutrino masses
$\,m_\nu^{}=\mathcal{O}(0.1\text{eV})\,$
at TeV scale [cf.\ Fig.\,\ref{fig:1} and Eq.\eqref{numassform2}].
Finally, in Section\,\ref{sec:5}, we studied the realization of a natural thermal or inflationary
leptogenesis through decays of the lightest singlet scalar $\sigma^{}_{1I}$
at a high scale around $\mathcal{O}(10^{13})$GeV.

 \vspace*{5mm}
\noindent
{\bf\large Acknowledgements}
\\[1mm]
We thank Alessandro Strumia for discussing the minimal dark matter models.
PHG was supported by the National Natural Science Foundation of China under Grant No.\, 11675100 and the Recruitment Program for Young Professionals under Grant No.\, 15Z127060004.  
HJH was supported in part by the National NSF of China (under grants 11675086 and 11835005)
and the National Key R\,\&\,D Project of China (under grant 2017YFA0402204);
he was also supported in part by the Shanghai Laboratory for Particle Physics and Cosmology
(under grant 11DZ2260700), and the Office of Science and Technology,
Shanghai Municipal Government (under grant 16DZ2260200).


\baselineskip 17pt

\vspace{5mm}
%

\end{document}